# Information retrieval in Current Research Information Systems.

**Andrei S. Lopatenko**
Vienna University of Technology
Gusshausstrasse 28 / E015
A-1040 Vienna, Austria
+43(1)58801-41573
andrei@derpi.tuwien.ac.at

**ABSTRACT**
In this paper we describe the requirements for research information systems and problems which arise in the development of such system. Here is shown which problems could be solved by using of knowledge markup technologies. Ontology for Research Information System offered. Architecture for collecting research data and providing access to it is described.

**INTRODUCTION**
Research data such as information about research results, projects, publications, organizations, researchers published on the web play more and more pervasive role in modern research. High dependence of modern research on already achieved research results produce requirements for research to have ability to retrieve research information in efficient way.

Information overloading, exponential rise of amount of information makes it difficult for researcher to find relevant information. To solve these problems a number of Current Research Information Systems (CRIS) is being developed.

But in most cases such system do not solve task of providing to researcher complete and actual information with minimum information noise. Researchers are not prone to publish results about their research in information systems, publishing usually limited to researcher's or project's homepages.

To provide actual and complete information for interested persons, information from research pages also should be included into information retrieval operations.

Usually researchers' or policy-makers' demands for research information is not limited to only information stored in any one the systems. Research information in any science or technology area is scattered among a number of heterogeneous information system. There is a strong need to gather information according request when it possible or to point researcher to systems where information can be found. It is very important to know if the gathered research information is actual and complete.

Experience of development university research CRIS in Finland[Lait-2000], ERGO project[ERGO] sponsored by European commission, showed that integration of data of research organizations is hardly to solve problem. Especially, if organizations are governed by different bodies or do not have direct benefits from participation in such networks.

So, it a necessity to find a solution for a problem data integration, which will be

- easy to implement for any participator
- flexible enough to embrace diversity and data meaning and structure in different organizations, sectors of science and states
- powerful to go provide sophisticated information retrieval services for users

We are developing information system (AURIS-MM Austrian Research Information System- MultiMedia enhanced) to provide research information of Austrian universities to interested consumers. The system is being developed as a substitution to AURIS (Austrian Research Information System) and FODOK (Research Documentation of Vienna University of Technology).

The new version of AURIS-MM is based on Semantic Web technologies

    RDF – Resource Description Framework www.w3.org/rdf

    RDFS – Resource Description Framework Schema www.w3.org/rdf

    DAML + OIL (DARPA Agent Markup Language + Ontology Inference Layer) www.daml.org

The results of the project were reported at the Workshop on Knowledge Markup and Semantic Annotation at K-CAP'2001 (Oct. 2001, Victoria)[Lop-KM] and EuroCRIS (European Current Research Information System) platform meeting(May 2001, Amsterdam)[EiuroCRIS9]

**ONTOLOGY DEVELOPMENT FOR SCIENCE**

The access to research information is one of the tasks important to researchers and some efforts already were done to provide to researchers, industry, policy-makers efficient information access to research data in some sectors of science, or to research limited to organization (university research information systems), or geographical boundaries (national networks, ERGO[ERGO] – European Research Information System) [Kul-2001].

The development and use of such system shown that it is almost impossible to collect complete data about research in sector or organization like an university.

Despite the huge amount of data scattered on internet web pages of projects, researchers, universities, it is hard to get researchers provide their data into centralized system.

Event if the universities holds actual or complete data, integration of databases is time and person-consuming tasks.

1. The most databases developed in different technologies (Relation DBMS, Oracle, MySql, XML-SGML)
2. The databases always have different structure and sometimes operate with different data. The databases should be harmonized
3. Database technologies such as replication require full transfer of data, what is often impossible due to political reasons.

In some countries (Denmark) national networks were created on Z39.50 technologies, but such technologies (Z3.50, LDAP) require use of some application profile by all participators. What is impossible due to diversity of database structure.

Full text search systems like Google (http://www.google.com) index among others also pages with research information. But they can not limit search to trusted data, understand context of the page and provide search based on meaning of the data.

On of the possible ways to collect data about research is a page annotation. Knowledge can be annotated on the page in a such way that automatic tools can collect and understand it [BL-2001, Hend-2001, Erd-2001]

Some efforts already done to develop mark up for scientific data.

Several ontologies were developed on SHOE (Simple HTML Ontology Extension). [Hefl-99, SHOE]

SHOE is a small extension to HTML which allows to annotate some knowledge about web page content. SHOW is very simple language for declaring ontology, defining classification, relationship, inference rules, categories, etc. SHOE was developed in Department of Computer Science, University of Maryland. SHOE specification, tools, SHOE ontology in plain text and DAML, examples are accessible at SHOE home page

Several ontologies for university and research data were developed for SHOE. There are University ontology and Computer Science Department ontology.

OIL(Ontology Inference Layer) [OIL, Fens-2000] - "is a proposal for a web-based representation and inference layer for ontologies, which combines the widely used modeling primitives from frame-based languages with the formal semantics and reasoning services provided by description logics. It is compatible with RDF Schema (RDFS), and includes a precise semantics for describing term meanings (and thus also for describing implied information)." OIL was sponsored by European Community via the IST projects Ibrow and On-To-Knowledge.

In the OIL for research data there were developed SWRC (Semantic Web Research Community Ontology) and KA2 Ontology of Knowledge Acquisition community

DAML (DARPA Agent Markup Language)[DAML] - ontology markup language, developed as an extension to RDF and RDFS. DAML allows specify ontologies and markup pages for automatic knowledge extraction. The last version of DAML is named DAML + OIL. DAML specifications, examples, tools, ontologies are published at DAML home page.

Several ontologies for research information are developed in DAML. Among them: DAML version of SHOE University ontology (http://www.cs.umd.edu/projects/plus/DAML/onts/univ1.0.daml), SWRC (Semantic Web Research Community) ontology (http://www.semanticweb.org/ontologies/swrc-onto-2000-09-10.daml), homework assignment ontology (http://www.ksl.stanford.edu/projects/DAML/ksl-daml-desc.daml).

More complete list of ontologies for research data as well metadata standards, thesaurus and system architectures at European Research Information Systems homepage (http://www.eurocris.org), Andrei Lopatenko's Resourse Guide to Metadata for Science, Research and Technology (http://derpi.tuwien.ac.at/~andrei/Metadata_Science.htm)

**ONTOLOGY**

So, the main of our ontology development was to develop ontology which will help persons interesting in research information retrieve relevant information.

Primary use cases of information retrieval for RIS are [Jeff-98, CERIF-2000, Lind-2000, Aks-2000]

- Retrieving information about research results by researchers or students for results reuse. The estimation of research results.
- Seeking collaborators which can take part in research projects as partners, sell their expertise, results and intellectual rights

- Finding facilities and equipment which can be used for research
- Assess to Research and Development capabilities by policymakers
- Finding ongoing research and technology activities and results of projects by users in commerce and industry

The ontology should contains terms already known to developers of Current Research Information system to make it more easy integrate new infrastructure with old ones

There are not a lot of metadata standard for science. The review of them have been done at [Grot-98,Lop-01].

Math-Net developed metadata format based on Dublin Core and RDF Schema for mark up of knowledge about content of researchers and institutes pages[MathNet]. Math-Net metadata set allows describe Researchers/Research groups/organizations, projects, results, events, publications.

In our ontology development we decided to use CERIF-2000 metadata standard (Common European Research Information Format)[CERIF-2000]. The first version of RDF format for science based on CERIF were already tested and reported[Ser-2001]

According to CERIF documents [CERIF] "CERIF 2000 is a set of guidelines meant for everyone dealing with research information systems. The CERIF 2000 guidelines are developed by a group of experts from the EU Member States and Associated Member states, under the co-ordination of the European Commission."

Now CERIF is used by several groups of developers and researcher in different EU states, it is proved and stable. Also different group of developers are well-acquainted with CERIF-2000 what will let make a process of ontology more easy

Despite excellence of CERIF as metadata format for research, there are certain lack in CERIF in description some types of research information resources. In development of our ontology we decided to enrich it with terms, slots from some other ontologies, to make it more suitable for research information retrieval.

In the next table is provided comparison of enriched CERIF ontology with a few already developed ontologies (they were described earler)

**Table 1. Comparison of selected ontologies for science**

| CERIF 2000 | Math-Net ontology | SWRC Semantic Web Research Community | University Ontology |
|---|---|---|---|
| Person Not classified in CERIF | Yes | Developed hierarchy suitable for research and education | Developed hierarchy suitable for research and education |
| Project Not classified in CERIF | Yes | Yes. Classified. | No |
| Organization | Yes | Close to CERIF classification | Only educational |
| Publication | Yes | Close to CERIF classification of publications. Grey literature is not included | Close to CERIF classification of publications. Grey literature is not included |
| Event | Conferences | Yes. Very close to CERIF | Conference |
| Equipment | No | No | No |
| Patent | No | No | No |
| Product | Only software and software libraries | Yes | Only software product |
| Expertise skill | Yes Subject Value | Research Topic | No |
| Multimedia elements No | No | No | No |
| Sites/pages No | Yes | No | No |

After the comparative analysis of the CERIF ontology, selected ontologies and some research information systems, it was recognized that CERIF ontology could be a base technology due to richness of base terms and relevance to RIS. But in some areas there are certain lacks in CERIF. Enriching of CERIF ontology with a terms from other can be useful for research information systems

The primitive units of the CERIF ontology are *Person*, *Project*, *Organization Unit*, *Publication*, *Event*, *Site* (Internet service/page), *Equipment*, *Result*, *Multimedia element*, *Research topic* (Expertise skill).

*Research results* which can be reused might be described in *publications* (articles, thesis, technical reports, etc.). *Research results* might be described precisely (*Research result* or *Product*). They can be presented by advanced presentation techniques - *Multimedia element*, which maybe video, images, drawing, diagrams, MS PowerPoint presentations.

*Research results* are results of research *projects,* invented by *persons(researchers, students),* in *organization units* (universities, labs, institutes, departments). Information about *expertise skills* of persons, organizations can be also significant for estimation of research results.

Some research results are patented and valuable information about them can be contained in *patents.*

To make search of research results more easy information about any entity can be classified by *research topics.*

To find a partner. Partner might be an *organization unit* or *person,* which has relevant for partner seeker *research results* and *experience.* Information about results and experience of partner can be extracted from its *publications*, description of the *projects.*

Information about organization units, publications, results, projects, persons can be stored on the *sites.* Of course, no one research information system can not store all relevant information and users need to know about other information system, which can help in search research results, partners.

To help user find information, data about other research data relevant *sites* and internet services should be provided to user.

Research may need *equipment* or *facilities.* Information about those entities also should be retrievable and searchable.

Table 2. Research Information Ontology terms

Organization unit
- Enterprise
- Higher Education Establishment
  - University
  - Faculty
  - Institute
- International organization
- Joint Research Center
- Non-research private non-profit
- Non-research public sector
- Private research center
- Private non-profit research center
- Public research center
- Laboratory
- Research Group

Project
- European project
- Fundamental research project
- Applied research project
- Financed by official bodies project

Person
- Researcher
- Student

Product/Research result
- Fundamental
- Applied
- Software
  - Software library
  - Information system
- Compound
- Process
- Technology
- Algorithm
- Documentation
  - Proposal

Event
- Conference
- Cultural event
- Exhibition
- Political event
- Sport event
- Trade fair
- Workshop

Publication
- Abstract
- Book
- Conference paper
- Conference proceedings
- Dissertation
- Guideline
- Index
- Journal article
- Lecture
- Multimedia

   Patent

   Report

   Review

Equipment

Multimedia element

   Audio

   AudioVisual

   DataForMultimedia

   ExecutableFile

   Flash

   Image

   RealMedia

   ShockWave

   Slide presentation

   Video

Site

   Organization's site

   Project's site

   Personal home page

   Publication on the web

   List of the publications

   Reference page

   Information system

      Library (access to articles)

      Research Information System (access to research data- projects, persons, organizations)

The complete ontology and set of terms are presented at http://derpi.tuwien.ac.at/~andrei/Metadata_Science.htm.

For ontology development CERIF-2000 Guidelines and Subject Index recommendations were used, as well Multimedia Ontology [Hunt-2001] and science and university ontologies mentioned early.

As a guidelines for ontology development we used [Noy-2001, Noy-G]

**INFORMATIONAL RETRIEVAL ARCHITECTURE**

The research data for retrieval should be collected, analyzed. To make possible analysis and understanding of meaning of data by software, they should be published in format understandable by software agent or annotated. Then annotations should be collected, analyzed, if it is considered necessary, they should also be transformed into one model/format. During search operation queries and data should be processed by search engines and response should be send to information consumers

So the process of information retrieval consists of

1. knowledge markup (by researcher)
2. harvesting marked-up knowledge by crawlers or software agents
3. transforming harvested data into formats appropriate for metadata repository/search engines
4. loaded into repository
5. retrieved by search engines according to users request

**WEB PAGE ANNOTATION**

So the ontology can serve for understanding meaning of data. But to make data understandable by software agents, they should be provided in a format, which agent can parse

A number of annotation tools are described in [Staab-2001].

For page annotation we use two tools: OntoMat and AURIS-MM metadata generating facilities.

OntoMat [OntoMat] is a user-friendly interactive webpage annotation tool. It includes web browser and ontology browser. Ontology browser supports DAML + OIL ontology exploration. Web browser supports web browsing, highlighting parts of the web pages and creating annotations based on highlighted part of the pages. To annotate the web page researcher needs to open web page in the browser, then open ontology from provided by project URL. Then the researcher can crate annotation highlighting regions of the page and describing them in ontology browser according to the ontology terms, relation and attributes. OntoMat automatically creates RDF annotation and new web page with included RDF annotation. The annotated web pages can be published on the web instead of annotated.

AURIS-MM metadata generating facilities generated RDF description of the data from AURIS-MM Relational database.

To create annotated web page, researcher needs input data about his research (projects, publications, etc) into AURIS-MM, and the use metadata generating facility just by pressing buttons. Generated RDF file then can be published on the web directly, or can be embedded into the web page.

The generated RDF file for the object has a persistent location in the AURIS-MM, which can be used as an identifier for that object. This is very important because information about the one object can be asserted on different pages. OntoMat supports only annotation and does not generate persistent URLs, because it is annotation tool.

Currently AURIS-MM does not support any ontology for semantic annotation as OntoMat does. But it supports vocabularies and thesaurus for advanced annotations, also it supports workflows and allows to re-use already inputted data.

**Fig. Annotation of the page**

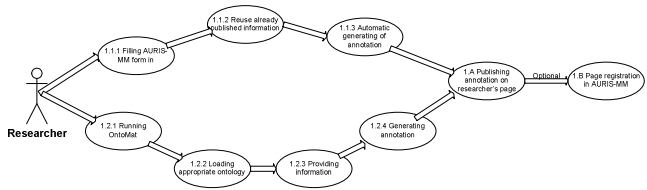

**Fig. The registration of multimedia element**.

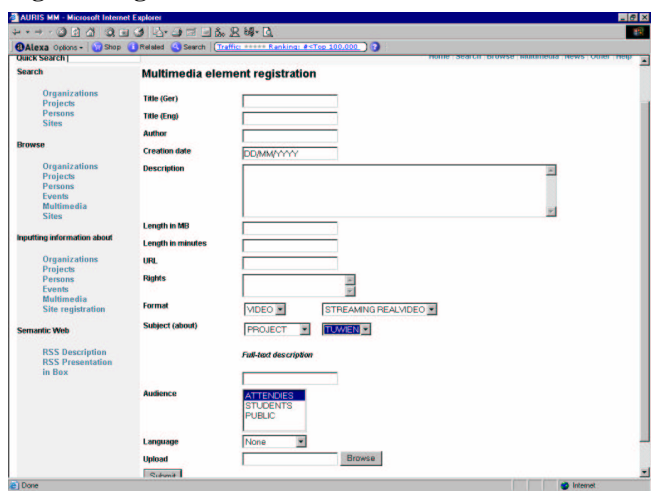

## COLLECTING METADATA

To make knowledge annotated on the web pages accessible for retrieval, it should be collected, analyzed, stored and made accessible for query engine.

Harvesting (collecting) RDF metadata possible by using RDF Crawler (http://ontobroker.semanticweb.org/rdfcrawl/index.html) – java application, which can crawl web pages and collect RDF data. After crawling RDF Crawler produces one file which store all RDF data and declaration of all used RDF Schemas.

**Fig. Metadata collecting into RDF database**

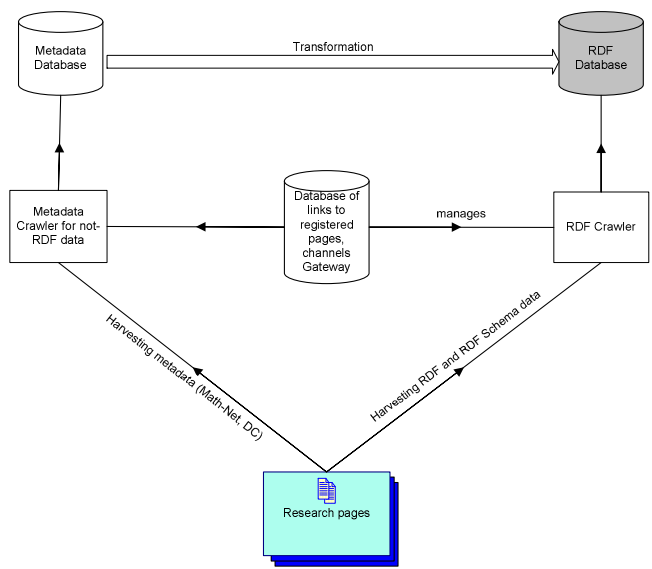

## QUERYING COLLECTED METADATA, GETTING KNOWLEDGE FROM ANNOTATIONS

Once the annotated metadata were collected how to use them.

There are several tools which can be used to search annotated pages.

SHOE Search Engine – Semantic Search (http://www.cs.umd.edu/projects/plus/SHOE/search/) search registered annotated pages. User of search engine can choose ontology, then choose type of resource he searches, create very simple filter conditions and search SHOE metadata database.

Our approach assumes that data would be described in RDF or can be translated into RDF by transformation procedure. Also to provide search services for researcher query facilities should be able to search data by its meaning (type of resource or property), values of attributes (properties) and relation between resources.

There are several query engines for RDF[Karv-2000], Squish, Ontobroker, Redland RDF Application Framework, MetaLog, RDF Data Query Language.

In our project to query RDF database Sesame RDF Query Repository and Querying Facility is used.

Sesame supports RQL (RDF Query Language) [Vass] which is being developed by ICS-FORTH Institute. Sesame supports storing both RDF and RDF Schema information. Querying Facilities of Sesame supports Schema information about subclasses and subproperties, searching by attributes values, resource relations.

OQL-like query engine supports which easily sea

Queryes

*http://derpi.tuwien.ac.at/~andrei/cerif.rdfs#Person*
All persons in database (and any subtype of a person,

| |
|---|
| -researchers and student) |
| *http://derpi.tuwien.ac.at/~andrei/cerif.rdfs#Researcher* |
| All persons who are researchers (or any subtype of researchers) |
| *^http://derpi.tuwien.ac.at/~andrei/cerif.rdfs#Researcher* |
| All persons, who are researchers and not any subtype of researcher |
| *select X,Y* |
| *from   #Project   {X}.   #project_persons{Y},   {Z} #expertise_skill {E}* |
| *where X = Z and N = "Semantic Web"* |
| All projects in Semantic Web with description of persons participation in them |
| If the organization or person, or Research Information System asserts new type of project – software project and in RDF Schema provides that it is a subtype of AURIS-MM, then it will also searched. |
| *select X,Y* |
| *from   ^#Project   {X}.   #project_persons{Y},   {Z} #expertise_skill {E}* |
| *where X = Z and N = "Semantic Web"* |
| Only projects in Semantic Web asserted as exactly CERIF projects and participants of those projects |

Sesame provides application interface through HTTP protocol, so application can query and update network RDF databases.

**CONCLUSIONS**

Use of Semantic Web technologies might be very fruitful for development of Research Information Systems.

The annotation of knowledge make it more easy to researchers and research organization to assert information about their research for dissemination. No need to register it in a number of information systems. Software agents can collect information and understand its meaning

Not only research data but also new domain knowledge can be also asserted and shared for use.

Semantic Web technologies solve a number of problems which are critical for implementation national-wide or European research information system They do not require approval of one format by all participant, using of the same vocabularies

Query engines for Semantic Web due to that inference abilities and schema exploration can make development of Research Information System more easy then conventional technologies like Relational Database management systems because exploration of domain knowledge is very crucial for CRIS systems .

**ACKNOWLEDGMENTS**

I thank Walter Niedermayer and all AURIS-MM project staff, Vienna University of Technology for support and helpful comments on previous versions of this article.